\documentclass[useAMS,usenatbib,usegraphicx]{mn2e}
\usepackage{journals}
\usepackage{times}
\usepackage[table]{xcolor}
\usepackage{ctable}
\usepackage{multirow}
\usepackage{natbib}
\usepackage{amsmath}
\usepackage{color}
\usepackage[utf8]{inputenc}
\usepackage{float}

\usepackage{graphicx}
\usepackage{caption}
\usepackage{subcaption}

\newcommand{\simgt}{\,\rlap{\lower 3.5 pt \hbox{$\mathchar \sim$}} \raise
1pt \hbox {$>$}\,}
\newcommand{\simlt}{\,\rlap{\lower 3.5 pt \hbox{$\mathchar \sim$}} \raise
1pt \hbox {$<$}\,}

\newcommand{\figwidth}{12cm}

\title{Dynamical signatures of infall around galaxy clusters: a generalized
  Jeans equation}
\author[Falco et al.]
{Martina  Falco$^1$,  Gary A. Mamon$^2$,
 Radoslaw Wojtak$^1$, Steen H. Hansen$^1$ 
\newauthor
and Stefan Gottl\"ober$^3$\\
$^1$ Dark Cosmology Centre, Niels Bohr Institute, University of Copenhagen,
Juliane Maries Vej 30, 2100 Copenhagen, Denmark\\
{\rm falco@dark-cosmology.dk},\\
$^2$ Institut d'Astrophysique de Paris (UMR 7095: CNRS \& UPMC),
  98 bis Bd Arago, 75014 Paris, France\\
$^3$ Leibniz-Institut f\"ur Astrophysik, An der Sternwarte 16,
  14482 Potsdam, Germany
}

\pubyear{2013}

\begin{document}

\maketitle

\begin{abstract}

We study the internal kinematics of galaxy clusters in the region beyond the sphere of virialization.
Galaxies around a virialized cluster are infalling towards the cluster centre with a non-zero mean radial velocity. 
We develop a new formalism for describing the dynamical state of clusters, by
generalizing the standard Jeans formalism with the inclusion of the
peculiar infall motions of galaxies and the Hubble expansion as well
as the contributions due to background cosmology.
Using empirical fits to the radial profiles of density, mean radial velocity
and velocity anisotropy of  both a stacked cluster-mass halo and two
isolated haloes
of  a cosmological dark matter only simulation,
we verify that our generalized Jeans equation correctly predicts the radial
velocity dispersion out to 4 virial radii.
We find that the  radial velocity dispersion inferred from the
standard Jeans equation is accurate up to 2 virial radii, but overestimated by $\approx 20\%$ for the
stacked halo and by $\approx 40\%$ for the isolated haloes, in the
range $\approx 2-3$ virial radii.
Our model depends on the logarithmic growth rate of the virial radius (function of
halo mass or concentration), which we
estimate in 7 different ways, and on the departure from
self-similarity of the evolution of the peculiar velocity profile in
virial units.
\end{abstract}


\section{Introduction}

Galaxy clusters are the largest gravitationally bound structures in the
Universe. Cluster studies represent a particularly deep source of information in modern
cosmology, since they provide constraints on the growth of structures in the
Universe and on cosmological parameters, in particular the dark energy
equation of state parameter $w$, from the evolution of the cluster mass
function \citep{HMH01,Voit05,CHF09}.  A crucial role is played by the
accuracy to which we can determine the cluster mass. Therefore, the
estimation of the cluster total mass has become an important research field,
which still remains a demanding task, mainly because most of the matter
content in clusters is not visible.

Clusters are characterized by a virialized region within which all components
(galaxies, intracluster medium and dark matter) are in rough dynamical
equilibrium, where galaxy motions are well described by the Jeans formalism.

The cluster mass distribution can be measured through many complementary
methods. The first approach to the cluster mass determination was the
application of the virial theorem to the member galaxies
\citep{Zwicky33}. More sophisticated techniques are based on the hydrostatic
measure of X-ray emissivity and temperature of the hot cluster gas
\citep{EFAJ02,Borgani+04,Zappacosta+06,SA07, HH11}, on the analysis of
large-scale velocity field \citep{MT05} and on the analysis of the galaxy
motions in clusters through the Jeans formalism. The radial profiles of total
mass and velocity anisotropy of clusters have been constrained by Jeans
analysis in several ways: predicting the observed radial profile of the
line-of-sight velocity dispersion \citep{Girardi+98}, as well as kurtosis
\citep{LM03, Lokas+06}; by isotropic \citep{KBM04} or anisotropic mass
inversion \citep{MB10}. Alternative methods to use galaxy motions, are by
fitting the $\Lambda$CDM distribution function \citep{Wojtak+08} to the
distribution of galaxies in projected phase-space density \citep{WLMG09,
  WL10} or by applying the caustic technique \citep{Diaferio99}.

Observations \citep{RD06}, N-body simulations \citep{MSSS04,WLGM05,CPKM08}
and a combination of both \citep{MMR11} have shown that virialized clusters
are surrounded by infall zones from which most galaxies move into the relaxed
cluster, as predicted by \cite{GG72}. These galaxies are gravitationally bound to the cluster but are not
fully virialized. This picture sparks multiple questions: does the infall
motion affect the standard formalism? Can we detect the effect of the infall
in cluster observations? Can this detection help to constrain the total mass
of clusters? Below we will attempt to answer some of these questions.

The dynamical and X-ray based mass estimators depend on the hypothesis that
the cluster is in steady-state dynamical or hydrostatic (for X-rays)
equilibrium. The presence of non-stationary motions just outside the virial
sphere may invalidate this assumption at that scale. For example, the
standard Jeans formalism involves outward integration along the
line of sight, beyond the virial radius, hence into the regions with negative
mean radial velocities, which are not accounted for.  Moreover, the
Jeans analysis relies on the assumption that the mean matter
density of the Universe does not contribute to the gravitational
potential (the so-called \emph{Jeans swindle}), and does not take into
account the effect of the expansion of the Universe.
Therefore, the mass
estimated through the usual methods may be significantly biased and not be
the true dynamical mass of the cluster. Two additional methods have been
developed to address this issue: gravitational lensing
(e.g. \citealp{MSBS10,Lombriser11}), and the caustic technique
\citep{Diaferio99,SDMB11}, which are both independent of the dynamical state
of the system. \cite{Zu2013} developed also a novel technique for constraining the radial profiles of the infall velocity from the projected velocity distributions of galaxies around clusters.
 
The aim of this paper is to generalize the standard Jeans formalism to
include radial streaming motions (i.e. infall), as well as cosmological terms.
This leads to a more general Jeans equation that also describes the outer
cluster region and simplifies to the standard Jeans equation when the infall
and cosmological corrections are negligible, like inside the virial radius.
Our motivation is to build a formalism with this \emph{generalized Jeans
  equation} to measure more accurately 
the mass profiles of clusters beyond the virial radius.
In Section~\ref{sec:theory}, we develop the formalism of the generalized Jeans
equation. We analyse a cosmological simulation in Section~\ref{sec:usesim} to
show that our generalized Jeans equation correctly reproduces the radial
velocity dispersion, and we determine the bias on the velocity
dispersion obtained with the standard Jeans equation.

\section{Non-equilibrium dynamics of galaxy clusters: the generalized Jeans equation}
\label{sec:theory}

The lowest-order Jeans equation relates the gravitational potential $\Phi$ of
the cluster to the dynamical properties of the galaxies. In spherical
coordinates, the \emph{standard Jeans equation} is (e.g., \citealp{BM82})
\begin{eqnarray}
\label{eqn:jeansstd}
-\rho(r)\frac{{\rm d}\Phi}{{\rm d}r} &=&-\rho(r)\frac{G M(r)}{r^2} \nonumber  \\
&=& \frac{{\rm d}(\rho\sigma_r^2)}{{\rm d}
  r}+2\,\frac{\beta}{r}\,\rho\sigma_r^2 \, . 
\end{eqnarray}
Here $\rho(r)$ is the density distribution of a tracer (e.g., the number
density of galaxies in and around clusters), $M(r)$ is the \emph{total} mass distribution (including dark
matter) in the cluster, $\sigma_r(r)$ is the galaxy velocity dispersion along
the radial direction and $\beta(r)$ is the velocity anisotropy parameter
defined by
\begin{equation}
\label{eqn:beta}
\beta(r)=1-\frac{\sigma_{\rm \theta}^2(r)+\sigma_{\rm \phi}^2(r)}{2 \sigma_r^2(r)}\, ,
\end{equation}
where $\sigma_{\rm \theta}(r)$ and $\sigma_{\rm \phi}(r)$ are the
longitudinal and azimuthal velocity dispersions (and are equal by spherical
symmetry).  
The anisotropy parameter
expresses the cluster's degree of radial velocity anisotropy. The value of
$\beta$ can vary from $\beta=-\infty$, corresponding to circular orbits
($\sigma_r=0$), to $\beta=1$, if orbits are perfectly radial ($\sigma_{\rm
  \theta}=\sigma_{\rm \phi}=0$). When $\sigma_{\rm \theta}=\sigma_{\rm
  \phi}=\sigma_{r}$ the system is isotropic ($\beta=0$).

In equation~(\ref{eqn:jeansstd}), we neglect streaming motions and any
time-dependence, i.e. the mean velocity components $\overline v_{\rm i}$ are
identically zero and, therefore, the velocity dispersions correspond to the
second moment of the velocity components $\sigma_{\rm i}^2=\overline{v_{\rm
    i}^2}$\, .

We now wish to go beyond the stationary approximation and to take into
account the possible presence of an infall motion of galaxies outside the
virialized core of clusters.

When we include the galaxies with mean radial velocity $\overline v_r\ne0$
and retaining time derivatives, the Jeans equation (obtained by taking the
first velocity moment of the collisionless Boltzmann equation) becomes
\begin{equation}
\label{eqn:jeansgen}
-\rho\frac{{\rm d}\Phi}{{\rm d}r}= \frac{\partial(\rho\overline{v_r^2})}{\partial
  r}+\frac{\rho}{r}\left[2\,\overline{v_r^2}-(\overline{v_{\rm
      \theta}^2}+\overline{v_{\rm \phi}^2})\right] 
+\frac{\partial(\rho\overline v_r)}{\partial t}\, .
\end{equation}   

The second-order velocity moment for the radial component is now 
  related to the radial velocity dispersion $\sigma_r$ and the first
  velocity moment, i.e. the mean radial velocity $\overline{v_r}$, by the general expression
\begin{equation}
\label{eqn:vs}
\overline{v_r^2}=\sigma_r^2+\overline{v_r}^2\, .
\end{equation}
We keep  $\overline{v_{\rm \theta}^2}=\sigma_{\rm \theta}^2$ and $\overline{v_{\rm \phi}^2}=\sigma_{\rm \phi}^2$, since we still assume no net longitudinal and azimuthal motions, i.e. we ignore bulk meridional circulation and rotation.
Using the continuity equation
\begin{equation}
\frac{\partial\rho}{\partial t}=-\frac{\partial(\rho\overline v_r)}{\partial r}-\frac{2}{r}\,\rho\overline v_r\, ,
\end{equation}
the Jeans equation~(\ref{eqn:jeansgen})  can be put in the following form:
\begin{eqnarray}
\label{eqn:jeansgen2}
-\rho\frac{{\rm d}\Phi}{{\rm d}r}&=&\frac{\partial(\rho\sigma_r^2)}{\partial
  r}+2\,\frac{\beta}{r}\rho\sigma_r^2 
+\rho \left[\overline v_r\frac{\partial \overline v_r}{\partial r}+\frac{\partial \overline v_r}{\partial t} \right]\, .
\end{eqnarray}
Therefore, the inclusion of non-equilibrated material leads to a modification
in the dynamical terms on the r.h.s. of the standard Jeans
equation~(\ref{eqn:jeansstd}), namely to the addition of two extra terms
involving the mean radial motion of galaxies. This correction is negligible
in the virialized core of the cluster, but it can become significant in the
outer region.

Since we are now considering distances very far from the centre of the cluster, we also need to take into account effects due to the underlying cosmology, meaning that the gravitational term in the Jeans equation also needs to be modified. 
The galaxies are subject to an attractive potential from the  mean density of the background and a repulsive potential from the cosmological constant, and when we add these contributions, the potential gradient is given by
\begin{equation}
\label{eqn:pot}
\frac{{\rm d}\Phi}{{\rm d}r}=\frac{G M(r)}{r^2}+\frac{4\pi}{3}G\rho_b r-\frac{1}{3}\Lambda r\, .
\end{equation}
Here $\rho_b$ is the mean density of the Universe, $\Lambda$ the cosmological constant and $H=\dot{a}/a$ the Hubble constant.
Introducing the dimensionless density parameter and cosmological constant commonly used
\begin{equation}
\Omega_{\rm m}=\frac{8\pi G\rho_b}{3H^2},  \quad  \Omega_\Lambda=\frac{\Lambda}{3H^2}\, ,
\end{equation}
equation~(\ref{eqn:pot}) reads
\begin{equation}
\label{eqn:pot2}
\frac{{\rm d}\Phi}{{\rm d}r}=\frac{G M(r)}{r^2}+q H^2 r\, ,
\end{equation}
 in terms of the deceleration parameter:
\begin{equation}
\label{eqn:q}
 q=-{\ddot{a}a \over \dot{a}^2}={\Omega_{\rm m}\over 2}-\Omega_\Lambda\, .
\end{equation}

In general, the radial velocity of galaxies can be written as the sum of the Hubble flow and a peculiar (infall) velocity flow:
\begin{equation}
\label{eqn:vel}
\overline v_r(r,t)=H(t)\,r+\overline v_{p}(r,t)\, ,
\end{equation}
and beyond the infall region surrounding the clusters, the peculiar velocity becomes negligible compared to the Hubble expansion:
$$
\overline v_r(r,t)\approx H(t)\,r   \quad  \text{when}   \quad   r\to\infty\, .
$$

One can now compute the non-stationary terms in equation~(\ref{eqn:jeansgen2}) using equation~(\ref{eqn:vel})
\begin{equation}
\label{eqn:pecvel}
\overline v_r\frac{\partial \overline v_r}{\partial r}+\frac{\partial
  \overline v_r}{\partial t} = \overline v_{p}\frac{\partial\overline
  v_{p}}{\partial r}+H\left(\overline v_{p}+r\,\frac{\partial\overline
  v_{p}}{\partial r}\right) 
-q H^2 r+\frac{\partial\overline v_{p}}{\partial t} \, ,
\end{equation}
where we have written the time derivative of the Hubble parameter in terms of equation~(\ref{eqn:q}) :
\begin{equation}
\dot{H}=-(q+1)\,H^2\, .
\end{equation}
Inserting equations~(\ref{eqn:pot2}) and (\ref{eqn:pecvel}) into
equation~(\ref{eqn:jeansgen2}), we obtain the \emph{generalized Jeans equation}
\begin{equation}
\label{eqn:jeansgen3}
\frac{\partial(\rho\sigma_r^2)}{\partial r}+2\,\frac{\beta}{r}\rho\sigma_r^2=-\rho\left[\frac{G M(r)}{r^2}+S(r,t)\right]\, .
\end{equation}

Here $\rho$
is the density of tracer. In this work, when applying the generalized Jeans equation
to test haloes, we will consider only the particles belonging to the
selected halo as our tracer. The mass $M(r)$ is thus the test halo
mass, without the contribution from other haloes or the diffuse Universe.

Equation~(\ref{eqn:jeansgen3}) differs from equation~(\ref{eqn:jeansstd}) by the inclusion of the new term
\begin{eqnarray}
S(r,t) &=& q\,H^2 r + \left ( v_r {\partial v_r\over \partial r} + {\partial
  v_r\over \partial t} \right ) \\
&=&\overline v_{p}\frac{\partial\overline v_{p}}{\partial r} 
+H\left(\overline v_{p}+r\,\frac{\partial\overline v_{p}}{\partial r}\right)+\frac{\partial\overline v_{p}}{\partial t} \, .
\label{eqn:Aofrt}
\end{eqnarray}

Equations~(\ref{eqn:jeansgen3}) and (\ref{eqn:Aofrt}) extend the standard
Jeans formalism to describe also the non-stationary dynamics of clusters, and
in principle, hold at any radius.

In equation~(\ref{eqn:Aofrt}), the background density and the cosmological constant contributions to the gravitational
potential, cancel exactly with the velocity term
relative to the pure Hubble flow. Thus, including all the effects
due to
the underlying
cosmology corresponds to applying the Jeans swindle \citep{FHWM12}.
 
Therefore, the extra term
$S(r,t)$ differs from zero only in the presence of infall velocity, i.e. setting $v_p(r,t)=0$, 
we immediately recover the standard Jeans
equation~(\ref{eqn:jeansstd}).

The most general solution of  equation~(\ref{eqn:jeansgen3}) provides the following expression for the radial velocity dispersion profile, depending on $\rho(r)$, $M(r)$, $\beta(r)$ and $\overline v_{p}(r,t)$ 
\begin{eqnarray}
\label{eqn:dispersion}
\sigma_r^2(r) &=&\frac{1}{\rho(r)} \,\exp\left[-2\int_0^r
\frac{\beta(s)}{s}\,{\rm d}s\right]
\nonumber \\ 
&\mbox{}& \times 
\,
\int_r^\infty \exp\left[2\int_0^{s}
\frac{\beta(\tilde{s})}{\tilde{s}}\,{\rm d}\tilde{s}\right]
\nonumber \\
&\mbox{}&  \qquad \quad \times \,\rho(s) 
\,\left[ \frac{G M(s)}{s^2}+S(s,t)\right]{\rm d}s\, ,
\end{eqnarray}
using equation~(\ref{eqn:Aofrt}) for $S(s,t)$.

\section{Comparison with cosmological simulations}
\label{sec:usesim}
\subsection{The simulation}

We analyse an N-body simulation with Wilkinson Microwave Anisotropy
Probe 3 (WMAP3) cosmological parameters, 
$\Omega_{\rm m}=0.24$, ${\Omega_\Lambda}=0.76$, the dimensionless
Hubble parameter $h=0.73$, the spectral index of primordial density
perturbations $n=0.96$ and the power spectrum normalization
$\sigma_{8}=0.76$. A box of size $160\,h^{-1}$Mpc and $1024^{3}$ particles was
used. Starting from a redshift $z=30$, the evolution was followed
using the message passing interface (MPI) version of the Adaptive Refinement Tree (ART) code. A hierarchical
friends-of-friends (FOF) algorithm was used for identifying clusters. The
linking length was $b=0.17$ times the mean inter-particle distance, corresponding
roughly to an overdensity relative to the mean of the Universe of $\Delta = (1.96/b^3)\,
[\ln(c+1)-c/(c+1)]\,(1+c)^2/c^2 \simeq 549\,(c/5)^{0.39}/(b/0.17)^3$
(adapted from \citealp{MKDG11}), where $c$ is the concentration parameter,
for haloes with a NFW density profile \citep{NFW96}.\footnote{The approximation $\Delta \simeq
  549\,(c/5)^{0.39}/(b/0.17)^3$ is accurate to 0.7\% for $2.5 < c < 10$.}
We define the virial radius of our haloes as the radius of overdensity
$\Delta = 387$ (i.e. overdensity of $\Delta_{\rm c}=92.8$ relative to
the critical density of the Universe) appropriate for the cosmology of the
simulation (with the approximation of \citealp{BN98}).

In order to test the generalized Jeans equation, which includes the effect of
the infall motion, we shall look at cosmological simulations of clusters, and
we demonstrate how to reproduce their radial velocity dispersions for radii
larger than the virial radius. To this end, we first need to choose functions
to parametrize the density, the mass, the anisotropy parameter and the infall
velocity of the simulation to handle them as analytical functions in
equations~(\ref{eqn:Aofrt}) and (\ref{eqn:dispersion}).  

\subsection{Analytical approximations to density, anisotropy and mean radial
  velocity profiles} 
\label{Analapprox}
N-body simulations show that the density distribution of a
dark matter halo, in the inner virialized region, is well described
by a double power-law profile 
\footnote{In the case of pure
dark matter haloes, the tracer corresponds to dark matter particles; thus,
$\rho$ in equation~(\ref{eqn:dispersion}) corresponds to the density
in equation~(\ref{eqn:density}).} \citep{Kra1998}
 \begin{equation}
\label{eqn:density}
\rho_{\rm h}(r)=\frac{\rho_{s}}{\left(r/r_{\rm s}\right)^{\eta}\left(1+r/r_{\rm s}\right)^{\xi}} \, .
\end{equation}
Here 
$r_{\rm s}$ is the scale radius, $\rho_{s}$ is the scale
density, $\eta$ and $\xi$ are the slopes, with values close to 1 and 2 respectively, which
correspond to the NFW model.

According to cosmological dark matter simulations (e.g. \citealp{ML05b} and
references therein), the radial anisotropy
typically varies from ($\approx 0-0.1$) at $r\approx 0$, increasing
with the distance and reaching a maximum value ($\approx 0.3-0.8$)
around one to two times the virial radius. Looking at larger radii, $\beta(r)$  also shows an almost universal trend: it drops to negative values, reaching a minimum, and then it approaches zero asymptotically \citep{WLGM05,AG08}.
If we define $r_{0}$ as the radius at which $\beta(r)$ passes through zero before becoming negative, we can parametrize the anisotropy function as
\begin{equation}
\label{eqn:betafunc}
\beta(r)=A\left(\frac{r}{r_{\rm v}}\right)^{\mu}\left(\frac{r_{0}-r}{r_{\rm v}}\right)\left[1+B\Bigl(\frac{r}{r_{\rm v}}\Bigr)^{\nu}\right]^{-\chi}\, ,
\end{equation}
where $r_{\rm v}$ is the virial radius.

\begin{figure}
\centering
\includegraphics[width=\hsize]{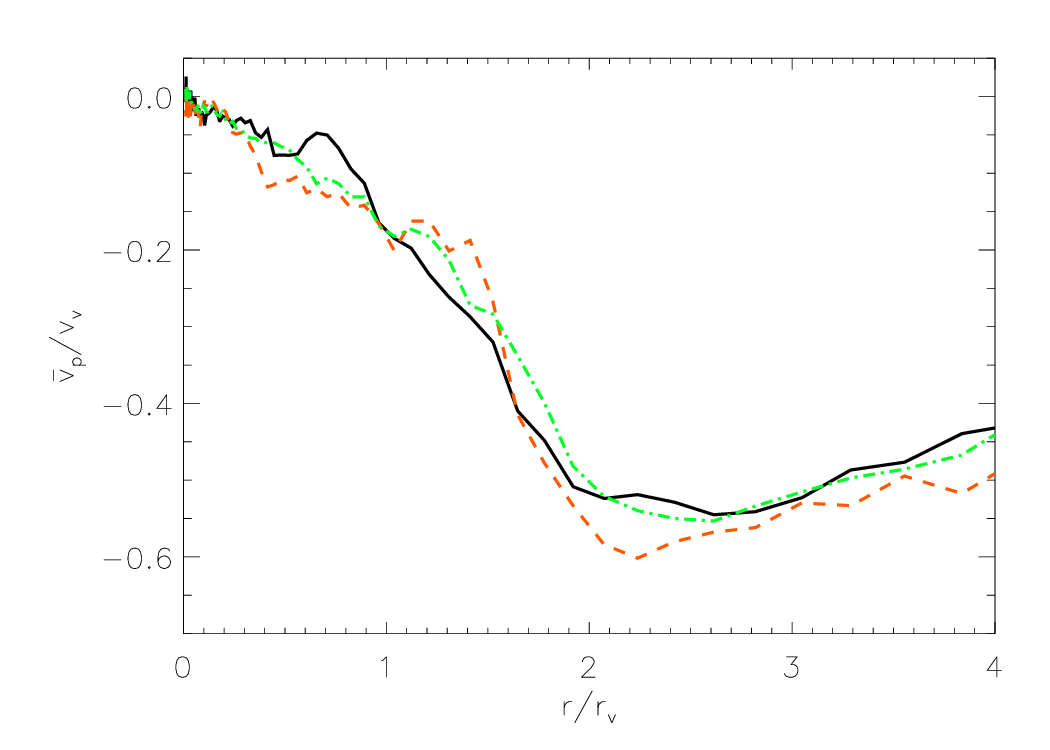}
	\caption{Mean peculiar radial velocity of three samples of stacked
      haloes. The mass ranges for the samples are: a very narrow mass
      range around $5\times 10^{13}\,
      M_{\odot}$
(green dot-dashed line), $(0.78-1.00)\times
      10^{14}\,  M_{\odot}$ (black, solid line) and $(1.00-1.27)\times 10^{14}\,
      M_{\odot}$ (red dashed line). 
}
\label{fig:3profiles}
\end{figure}

The new formalism also includes the mean infall velocity of galaxies as an
additional unknown function.  Simulations show a quite universal trend for
the radial profile of the mean velocity up to very large radii.
Fig.~\ref{fig:3profiles} displays the mean radial velocity with the Hubble
flow subtracted, i.e. the peculiar component $\overline v_{p}$, up to 4
virial radii, for three samples of stacked halos. 
The samples contain the same number of haloes and the mass ranges
  are: a very narrow bin around $5\times 10^{13}\,
M_{\odot}$ (green dot-dashed line), $(0.78-1.00)\times 10^{14}\,
M_{\odot}$ (black, solid line) and
$(1.4-1.8)\times 10^{14}\, M_{\odot}$ (red dashed line).  The velocity is negative everywhere,
clearly showing the infall motion, particularly pronounced between $\approx
1$ and $4\,r_{\rm v}$.  The three profiles appear to look very
similar.  

In the innermost region, the cluster is fully equilibrated ($\overline v_{\rm r}=0$). The peculiar velocity profile can then be approximated, for $r\ll r_{\rm v}$, as
 \begin{equation}
\label{eqn:limit}
\overline v_{p}(r,t) \approx-H\,r \quad
\, .
\end{equation}
In general,
the mean peculiar velocity can be written as
 \begin{equation}
\label{eqn:velocity}
\overline v_{p}(r,t)=-H(t)\,r_{\rm v}(t)\,f\left(\frac{r}{r_{\rm v}(t)}\right)\ ,
\end{equation}
where $f({r}/{r_{\rm v}})$ must be such that the condition ~(\ref{eqn:limit})
is satisfied. 
As we will show in the next sections, the function  $f\left({r}/{r_{\rm v}}\right)$
is well approximated by the formula
\begin{equation}
\label{eqn:velocity1}
f\left(\frac{r}{r_{\rm v}}\right)=\left\{\left[\left(\frac{r}{r_{\rm v}}\right)^{-a}+C\left(\frac{r}{r_{\rm v}}\right)^{b}\right]^{1/a}-D\right\}^{-1}\, .
\end{equation}
Equation~(\ref{eqn:Aofrt}) involves the time derivative of the
radial infall velocity.
Equations~(\ref{eqn:velocity}) and ~(\ref{eqn:velocity1}) describe a profile where the dependence on time is through
$H(t)$ and $r_{\rm v}(t)$. 
The function that describes the radial shape of the velocity,
equation~(\ref{eqn:velocity1}), might change in time as well. We
parametrize this dependence by multiplying
equation~(\ref{eqn:velocity}) by a factor that involves time only:
\begin{equation}
\label{eqn:velocitytime}
\overline v_{p}(r,t)=-H(t)\,r_{\rm v}(t)\,f\left(\frac{r}{r_{\rm v}(t)}\right)\,
\left ({t\over t_0}\right)^\alpha\ ,
\end{equation}
where $t_0$ is the present age of the Universe.

We can now explicitly calculate the time derivative of
$v_{p}(r)$ and complete the computation of the extra term $S(r,t)$, which
then obeys

\begin{eqnarray}
{t\,S(r,t)\over v_p(r,t)} &=&
H\,t\,\left [1+{{\rm d}\ln f\over {\rm d}\ln x}-{{\rm d}f\over {\rm
      d}x}\,\left ({t\over t_0}\right)^\alpha \right]
\nonumber \\
&\mbox{}& \quad 
+ {{\rm d}\ln H\over {\rm d}\ln t} 
+ {{\rm d}\ln r_{\rm v}\over {\rm d}\ln t} \,\left (1-{{\rm d}\ln f\over {\rm
    d}\ln x} \right) + \alpha \ ,
\label{eqn:Anew}
\end{eqnarray} 
where $x = r/r_{\rm v}$.

The parameter $\alpha$ describing the departure from self-similarity of the
evolution of the infall velocity profile in virial units is not well known.
In Appendix~\ref{sec:nonselfsim}, we analyse fig.~13c of \cite{CPKM08},
describing this evolution for stacked haloes, to
deduce that $\alpha \approx -0.55\pm0.1$ (see Fig,~\ref{fig:vpcuesta}).
Since we are analysing our simulation at $z=0$, 
the precise value of $\alpha$ probably depends on the radial shape of the velocity
at the present time. We
consider it as a free parameter of our model, and we expect it to vary
with a small scatter, when considering different velocity profiles.

In Appendix~\ref{sec:rvgrowth}, we estimate  ${\rm d}\ln r_{\rm v}/{\rm d}\ln t$ using the mass
growth rates measured in cosmological simulations as well as through
analytical theory, to conclude that the mean growth of haloes at
$z=0$ for the cosmology of our simulation and for the halo mean mass and
concentration parameter is ${\rm d}\ln r_{\rm v}/{\rm d}\ln t \simeq 0.7$.
We note that observers will tend to discard clusters having undergone recent
mergers, while we stack 27 haloes regardless of their recent merger history,
so that observers will effectively choose halos with slightly smaller values
of ${\rm d}\ln r_{\rm v}/{\rm d}\ln t$.
But in Appendix~\ref{sec:rvgrowth}, we also compute the minimal growth of a
fully isolated halo in an expanding universe, and find 
${\rm d}\ln r_{\rm v}/{\rm d}\ln t \simeq 0.68$, which is only very slightly lower.

Having expressions for $\rho$, $M$, $\beta$ and $\overline v$, we can then
use equation~(\ref{eqn:dispersion}) to calculate the radial velocity
dispersion of simulated haloes, which we can compare to the true dispersion profile.
In the next sections, we show the results obtained for a sample of
stacked haloes and for two isolated haloes.

\subsection{Comparison with a stacked halo}
\label{sec:stacked}

We begin by selecting a sample of 27 stacked cluster-size haloes from our
simulation, with virial masses in the range $[(0.78-1.0)\times10^{14}\,
\rm  M_\odot]$. We denote this sample as our \emph{stacked halo}.
The characteristic quantities of the stacked halo are taken as the mean of
the 27 individual haloes, and are listed in Table~\ref{tbl:halos}.

\begin{table}
\caption{Virial parameters and concentrations of the haloes.}
\begin{center}
\begin{tabular}{lcccc}
\hline
& $M_{\rm v}\ (10^{13}\,M_\odot)$ 
& $r_{\rm v}\ (\rm Mpc)$ 
& $v_{\rm v}\ (\rm km \, s^{-1})$ 
& $c$ \\
\hline
Stacked & 8.6 & 1.14 & 569 & 6.4 \\
Halo 1 & 8.4 & 1.13 & 566 & 6.9 \\
Halo 2 & 7.8 & 1.10 & 553 & 7.6 \\
\hline
\end{tabular}
\end{center}
\label{tbl:halos}
\end{table}

\begin{figure}
\centering
    \includegraphics[width=\hsize]{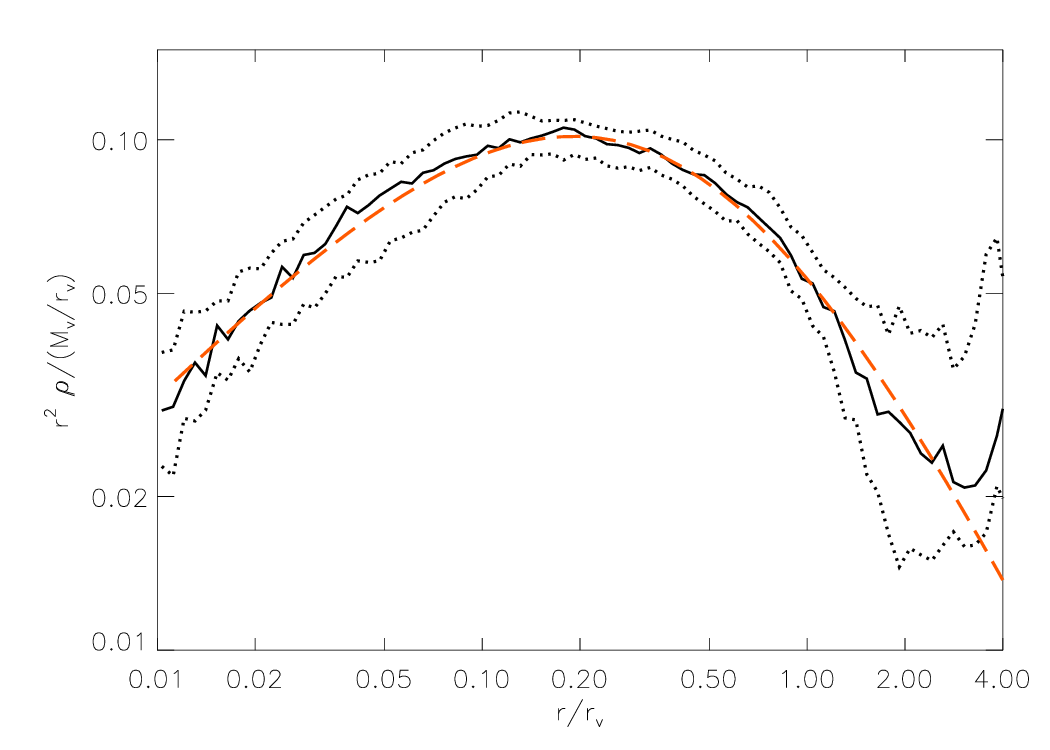}  
    \includegraphics[width=\hsize]{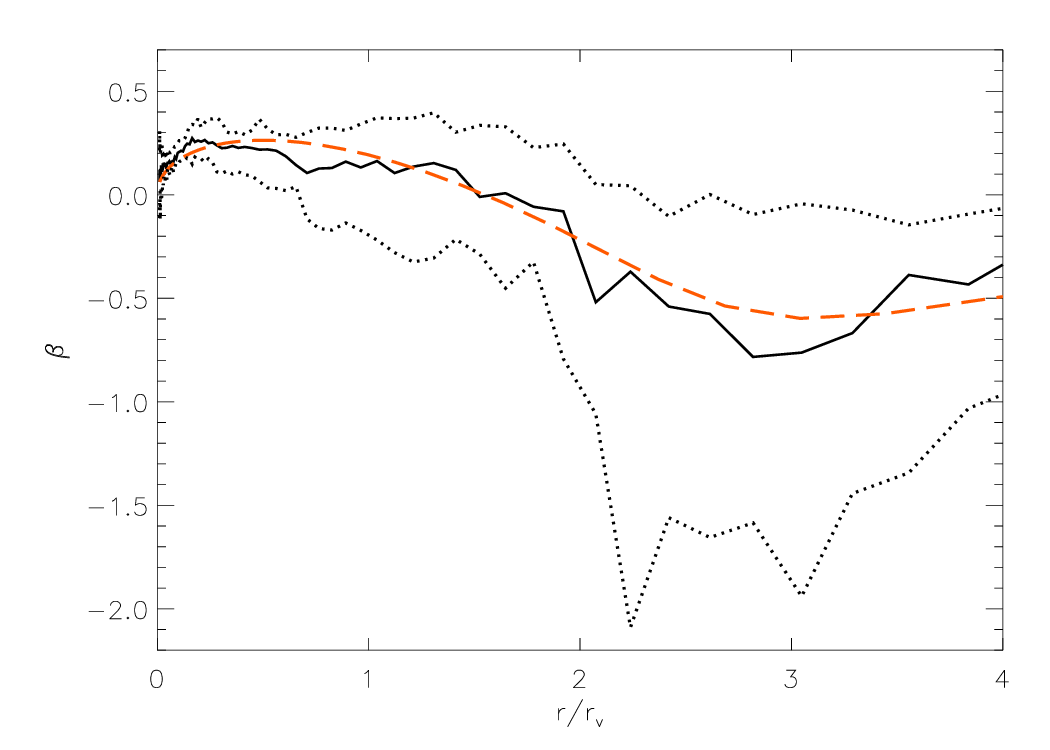}  
\caption{Density and anisotropy profiles of the stacked
       halo. Top panel: comparison between the quantity
       $r^{2}\rho(r)$, where $\rho(r)$ is the median density profile,
       from the simulation (black solid line), and from the parametric
       fit, with $\rho(r)$ given by equation (\ref{eqn:density}) and
       best parameters listed in Table~\ref{par2} (red dashed
       line). The black dotted lines denote the quartiles. Bottom panel: median (black solid) and quartiles
       (black dotted) velocity anisotropy profile. The red dashed line
       shows the fitting function (\ref{eqn:betafunc}) with parameters quoted in Table~\ref{par2}.}
\label{fig:densityanisotropy}
\end{figure}

\begin{figure}
\centering
\includegraphics[width=100mm,scale=0.2, bb =110 10 440 355,clip]{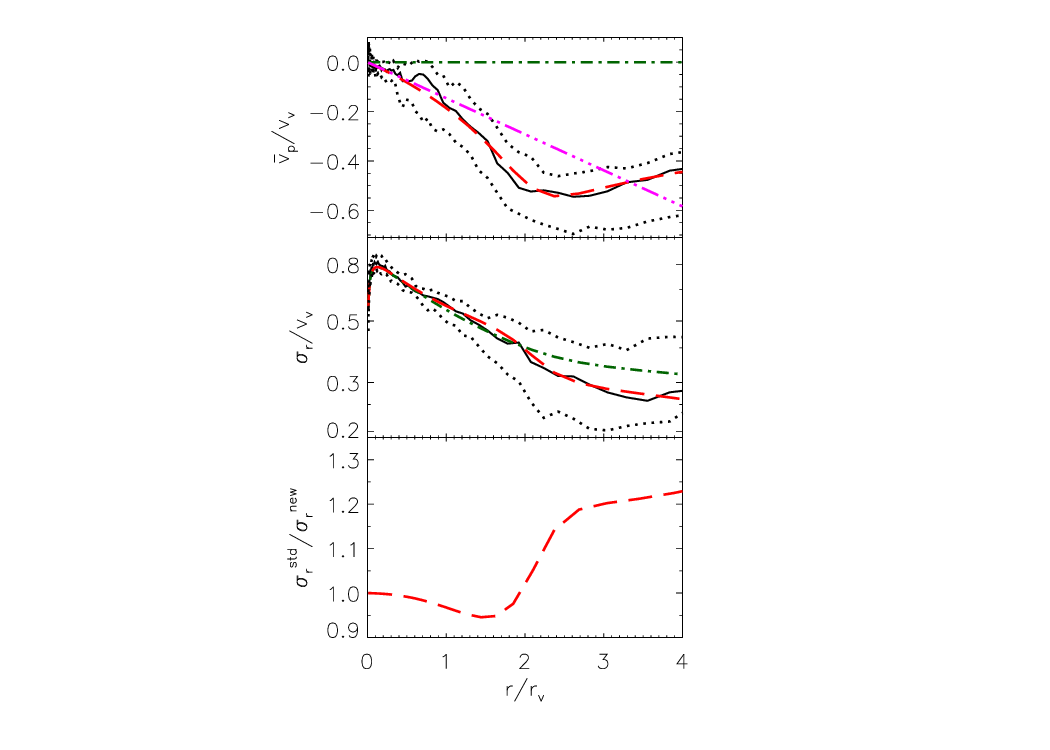}
	\caption{Mean peculiar radial velocity and radial velocity dispersion of
     the stacked halo. Top panel: 
median (solid black) and quartiles (dotted black) of the mean peculiar radial velocity profile  measured in the
simulation, and a fit to the form in equation~(\ref{eqn:velocity}) with parameters
  quoted in Table~\ref{par2} (red dashed line).
The green dash-dotted line represents the case of zero peculiar
velocity. The magenta triple-dot-dashed line corresponds to the
inner limit $\overline v_{p}\approx-H r$ for $r\ll r_{\rm v}$.
Middle panel: comparison between the radial velocity
      dispersion profile of the stacked halo, measured in spherical shells 
(black solid line),
and the one computed by
  equation~(\ref{eqn:dispersion}), where $\rho(r)$, $\beta(r)$ and
  $\overline v_{p}(r)$ are approximated with
  equations~(\ref{eqn:density}), (\ref{eqn:betafunc}) and (\ref{eqn:velocity}), with the
  parameters of Table~\ref{par2} (red dashed line). We set
  $\alpha=-1.0$ in equation~(\ref{eqn:Anew}).  
The green dash-dotted line shows the velocity dispersion profile from the
standard Jeans equation~(\ref{eqn:jeansstd}), 
i.e. for zero mean peculiar velocity. The black dotted lines represent the quartiles.
Bottom
  panel: ratio between the dispersion calculated with the standard
Jeans equation (the green dash-dotted line in the middle panel) and
with our new generalized equation (the red dashed line in the middle panel).
}
\label{fig:sigmainfsamplestack}
\end{figure}

Fig.~\ref{fig:densityanisotropy} shows the profiles of $r^2\rho(r)$ (top
panel) and $\beta(r)$ (bottom panel) for the stacked halo. The black
solid lines denote the median profiles from the simulation, the black dotted
lines correspond to the first and the third quartiles, and the red dashed
lines represent our fits.

The density can be well approximated in the
region ($r<3\,r_{\rm v}$) using
the double-power formula of
equation~(\ref{eqn:density}) , 
with parameters
listed in Table~\ref{par2}. The simulated density profile increases beyond a
turn-around radius of $3-4\,r_{\rm v}$,  due to the
presence of other structures surrounding the halo. Our model describes
an isolated system; therefore, we do not take into account the
presence of other structures.

The anisotropy profile is well fitted  by equation~(\ref{eqn:betafunc}) up
to $4\,r_{\rm v}$.
In our case, $r_{0}=1.55\,r_{\rm v}$, and the other best-fitting parameters for the
 anisotropy are listed in
 Table~\ref{par2}.

 According to equation~(\ref{eqn:dispersion}), the radial velocity dispersion
 also requires the knowledge of the mean peculiar radial velocity profile of
 the sample. 
In the upper panel of Fig.~\ref{fig:sigmainfsamplestack} are displayed 
the mean peculiar velocity of our sample (black solid line), quartiles (black
dotted lines), and $\overline v_{p}(r)$ given by formula
(\ref{eqn:velocity}), with parameters listed in Table~\ref{par2} (red dashed
line).
The green dash-dotted line represents the case of zero peculiar
velocity.

The radial velocity dispersion is shown in the central panel of
Fig.~\ref{fig:sigmainfsamplestack}. The black solid line shows the
simulated profile and the black dotted
lines correspond to the first and the third quartiles.
The green dash-dotted line shows the velocity dispersion profile from the
standard Jeans equation~(\ref{eqn:jeansstd}).
We compute the velocity dispersion by using
equation~(\ref{eqn:dispersion}) where the extra term is given
by equation~(\ref{eqn:Anew}) and best parameters in
Table~\ref{par2} (red dashed line). 
We find that the best solution
is given by setting $\alpha=-1$.
This value of $\alpha$ is more negative than the value of $\approx -0.55\pm0.1$ that
we infer (Fig.~\ref{fig:vpcuesta}) 
from the evolution of the stacked cluster-mass halo of \cite{CPKM08}.

\begin{table}
\caption{Parameters of empirical fits to $\rho(r)$ (with  eq.~[\ref{eqn:density}]), $\beta(r)$ (with
  eq.~[\ref{eqn:betafunc}]) and  $v_{p}(r)$ (with eq.~[\ref{eqn:velocity}]) for the stacked haloes.}
\label{par2}
\center
\begin{tabular}{ clclclc}
\hline
$\rho(r) $ & $\beta(r)$ & $v_{p}(r)$  \\
\hline
$\rho_s=2.4\,M_{\rm v}/r^3_{\rm v} $  & $\mu=0.48$ &      $a=9.7$      \\
 $r_s=0.38\,r_{\rm v}$ & $\nu=9.5$ &     $b=2.8$       \\
$\eta=1.35$ & $\chi=0.37$ &      $C=5.7\times\,10^{-5}$     \\
$\xi=1.96$ & $A=0.35$ &            $D=0.214$    \\
$$ & $B=5\times 10^{-5}$ &      $$ \\      
\hline                     
\end{tabular}
\end{table}

The radial velocity dispersion profile measured
for the halo matches very well the one predicted
by our generalized Jeans equation~(\ref{eqn:dispersion})  all the way out to $4\,r_{\rm v}$.
The standard
 Jeans formalism can predict $\sigma_r$ up to $\approx 2\,r_{\rm v}$, where
 the infall and the cosmological corrections are still very
 small or cancel out.
In the region where these contributions are significant the
 radial velocity dispersion inferred from
equation~(\ref{eqn:jeansstd}) is overestimated by $\approx 20\%$ in the
range $\approx 2-3$ virial radii, as shown in the bottom panel of
Fig.~\ref{fig:sigmainfsamplestack}. The slope of 
$\sigma_r$ becomes steeper in the region of non-equilibrium, and this is well reproduced by
 equation~(\ref{eqn:dispersion}). 

In Appendix~\ref{sec:slopes}, we also present the results of the simulations at slightly larger radii, and consider relations between parameters of interest, in particular we show the derivative of the density, $\gamma$, and of the radial velocity dispersion, $\kappa$. We also present plots in the two-dimensional spaces $\gamma-\kappa$ and $\gamma-\beta$ for distances up to $\approx 6 r_{\rm v}$.

\subsection{Comparison with isolated haloes}

The example in the previous section demonstrates the correctness of this generalized formalism when
applied to a stacked sample of cluster-size haloes. However, when
taking the median profiles of the stacked sample, we are not guaranteed that the
individual halos will have the same profiles.
Our median quantities based upon a sample with an odd number of haloes
effectively correspond to a single halo. 
However, the halo involved in the median density at a given radius, may not be the same as that involved in the median radial velocity, or that for the median velocity anisotropy.

A further test is therefore to apply the same approach also to
individual haloes in the sample, in order to be independent of the
analysis of the median profiles.
Our model describes the dynamics
of clusters when they are isolated. Thus, we need to look for haloes in
our sample which are not surrounded by massive  neighbours. In
particular, we have searched for haloes which have no neighbours
with mass at least half of theirs, within a distance of $10\, r_{\rm v}$.
With this criterion, we have selected two optimal haloes among the 27
belonging to our sample. We denote our haloes as \emph{halo~1} and
\emph{halo~2}. 
The virial masses, radii, velocities as well as their concentrations 
are listed in Table~\ref{tbl:halos}.

\begin{table}
\caption{Parameters of empirical fits to $\rho(r)$ (with  eq.~[\ref{eqn:density}]), $\beta(r)$ (with
  eq.~[\ref{eqn:betafunc}]) and  $v_{p}(r)$ (with
  eq.~[\ref{eqn:velocity}]) for \emph{halo 1} and \emph{halo 2}.}
\label{par3}
\center
\begin{tabular}{ clclclc}
\hline
$$ & $\rho(r) $ & $\beta(r)$ & $v_{p}(r)$  \\
\hline
\multirow{6}*{halo 1}   & $\rho_s=2.45\,M_{\rm v}/r^3_{\rm v} $  & $\mu_{\rm inner}=0.5$ &  $a=4.75$        \\
                                    & $r_s=0.4\,r_{\rm v}$ & $\mu_{\rm outer}=2.0$ &   $b=2.5$        \\
                                    & $\eta=1.37$ &$\nu=23.0$ &   $C=4.8\times\,10^{-3}$      \\
                                    & $\xi=2.22$ &$\chi=0.23$  &      $D=0.29$             \\
                                    & $$ &$A=0.6$ &      $$ \\      
                                    & $$ &  $B=6\times 10^{-11}$ & $$\\
\hline            
\multirow{6}*{halo 2}   & $\rho_s=41\,M_{\rm v}/r^3_{\rm v} $  & $\mu_{\rm inner}=0.57$ &   $a=4.8$      \\
                                    & $r_s=0.11\,r_{\rm v}$ & $\mu_{\rm outer}=2.0$ &  $b=2.6$    \\
                                    & $\eta=0.9$ &$\nu=24.6$ &        $C=2.4\times\,10^{-3}$   \\
                                    & $\xi=2.1$ &$\chi=0.25$  &    $D=0.21$       \\
                                    & $$ &$A=0.5$ &      $$ \\      
                                    & $$ &  $B=2.1\times 10^{-11}$ & $$ \\   
\hline                 
\end{tabular}
\end{table}

\begin{figure}
\centering
    \includegraphics[scale=0.5]{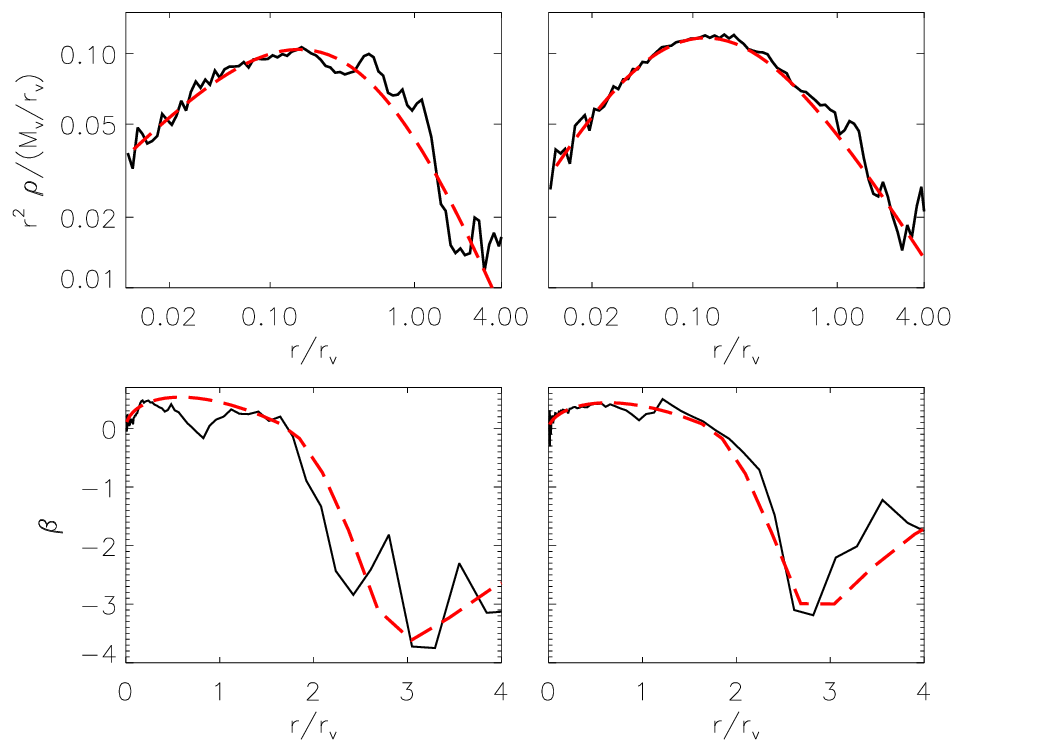}  
\caption{Density and anisotropy profiles of the isolated
       haloes. The left-hand panels correspond to \emph{halo 1} and the
       right-hand panels correspond to \emph{halo 2}. Top panels: comparison between the quantity
       $r^{2}\rho(r)$, where $\rho(r)$ is the radial density profile,
       from the simulation (black solid line), and from the parametric
       fit, with $\rho(r)$ given by equation (\ref{eqn:density}) and
       best parameters listed in Table~\ref{par3} (red dashed
       line). Bottom panels:
       radial velocity anisotropy profile (black solid line). The red dashed line
       shows the fitting function (\ref{eqn:betafunc}) with parameters quoted in Table~\ref{par3}.}
\label{fig:densityanisotropyH}
\end{figure}

Fig.~\ref{fig:densityanisotropyH} shows the profiles of $r^2\rho(r)$ (top
panels) and $\beta(r)$ (bottom panels) for the two individual
haloes. We fit the profiles with, respectively, equations~(\ref{eqn:density}) for the density and~(\ref{eqn:betafunc}) for the
anisotropy. In the outer region, where the anisotropy takes negative
values, the profiles of both haloes are much steeper
than the median profile of the stacked halo. Therefore, in
equation~(\ref{eqn:betafunc}) it is convenient to use two different slopes: $\mu_{\rm inner}$
for $r<r_{0}$ and $\mu_{\rm outer}$ for $r>r_{0}$. 
For both haloes,  $r_{0}=1.75\,r_{\rm v}$.
The best-fitting parameters
are shown in Table~\ref{par3}.

\begin{figure}
\centering
\includegraphics[width=87mm, height=70mm]{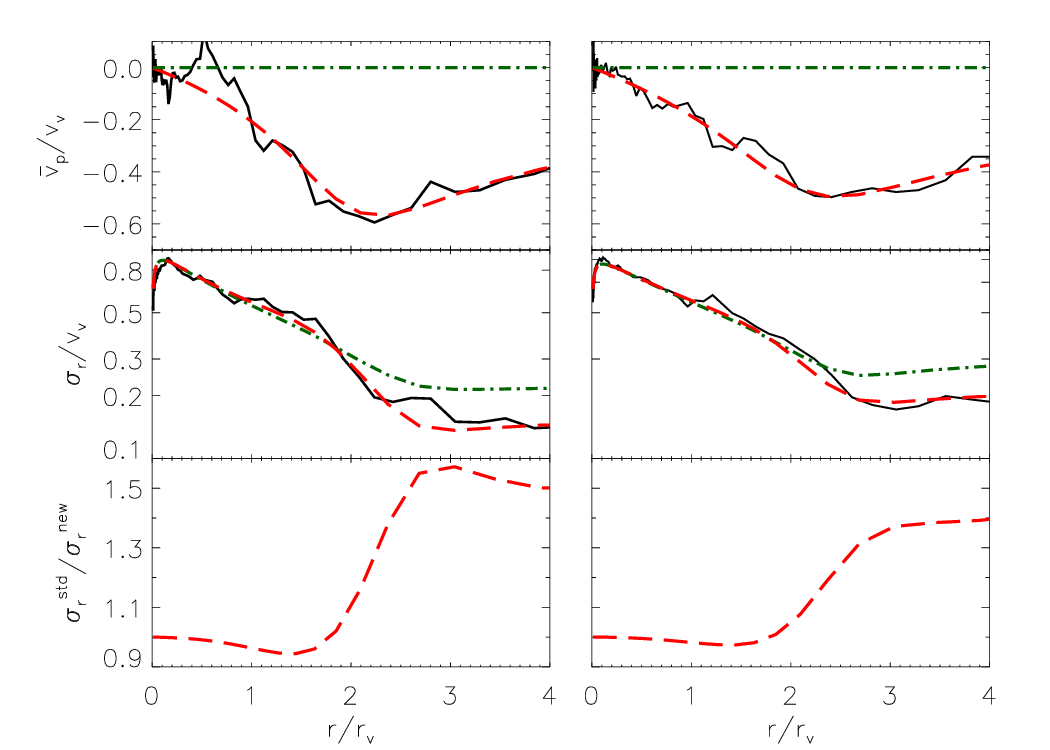}
	\caption{Mean peculiar radial velocity and radial velocity dispersion of
     the isolated haloes. The left-hand panels correspond to \emph{halo 1} and the
       right-hand panels correspond to \emph{halo 2}. Top panels: 
mean peculiar radial velocity profile  measured in the
simulation (black solid line), and a fit to the form in equation~(\ref{eqn:velocity}) with parameters
  quoted in Table~\ref{par3} (red dashed line).
The green dashed-dotted line represents the case of zero peculiar
velocity.
Middle panels: comparison between the radial velocity
      dispersion profile, measured in spherical shells 
(black solid line) 
and the ones computed by
  equation~(\ref{eqn:dispersion}), where $\rho(r)$, $\beta(r)$ and
  $\overline v_{p}(r)$ are approximated with
  equations~(\ref{eqn:density}), (\ref{eqn:betafunc}) and (\ref{eqn:velocity}), 
with the
  parameters of Table~\ref{par3} (red dashed line).  For both haloes, we set
  $\alpha=-0.65$ in equation~(\ref{eqn:Anew}). 
The green dash-dotted line corresponds to the velocity dispersion profile from the
standard Jeans equation~(\ref{eqn:jeansstd}), 
i.e. for zero mean peculiar velocity. 
Bottom
  panels: Ratio between the dispersion calculated with the standard
Jeans equation (the green dash-dotted line in the middle panel) and
with our new generalized equation (the red dashed line in the middle panel). 
}
\label{fig:sigmainfsampleH}
\end{figure}

The infall velocity profiles of the haloes are displayed in the upper
panels of Fig.~(\ref{fig:sigmainfsampleH}). 
The corresponding radial velocity dispersion profiles are shown in the middle panels. The black solid lines denote the simulated
profiles, and the red dashed lines correspond to the solution of
equation~(\ref{eqn:dispersion}), in the case of $v_{p}(r)$ given by our
fits. We find that the value $\alpha=-0.65$ provides a good match to
the velocity dispersion for
both haloes. The green dash-dotted lines denote the velocity dispersion profiles from the
standard Jeans equation~(\ref{eqn:jeansstd}).

Also in the case of these two isolated haloes, we find that our
prediction of the velocity dispersion matches very well the measured
profiles. The standard Jeans solution provides a good match up to
$\approx\,2\,r_{\rm v}$, while our generalized Jeans equation improves
the match at distances $\approx 2-4\,r_{\rm v}$.

The bottom panels of Fig.~\ref{fig:sigmainfsampleH} show the ratio between the standard Jeans
solution and our generalized solution. For the isolated haloes, the
ratio is slightly larger than the one computed for the stacked sample.
The velocity dispersion
calculated by the standard Jeans equation is overestimated by $\approx
40\%-60\%$ for the first halo and $\approx
40\%$ for the second halo,
in the
range $\approx 2-4$ virial radii. 

\begin{figure}
\centering
	\includegraphics[width=8.5\figwidth]{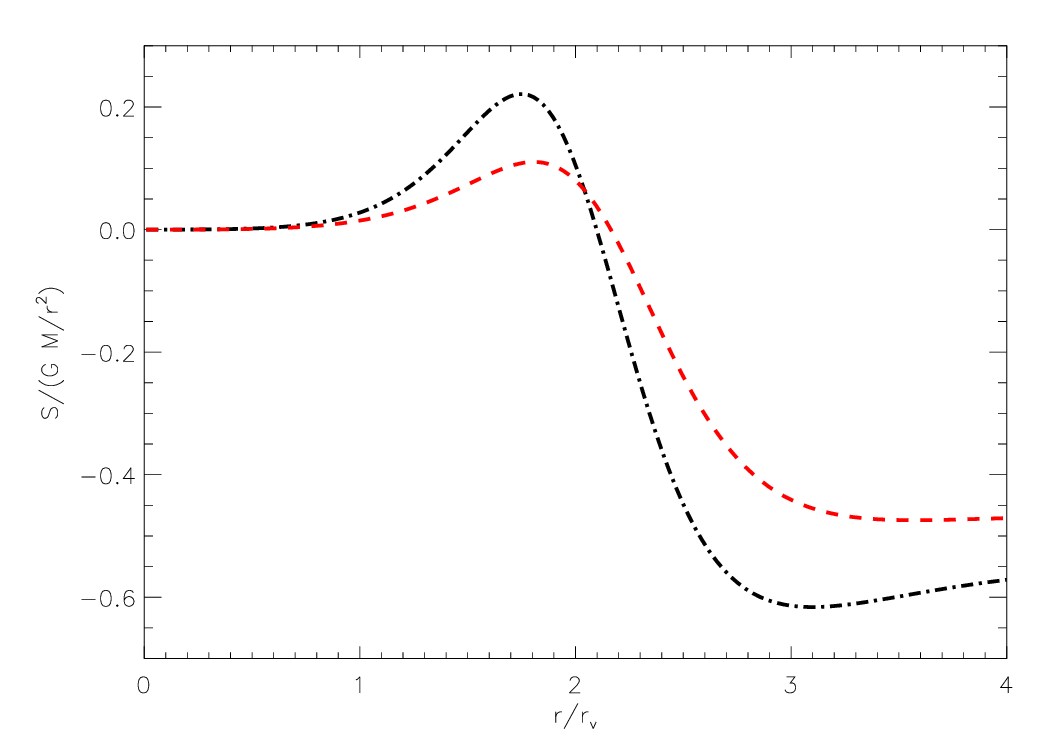}
	\caption{Relative mass excess deduced from the standard Jeans
      equation~(\ref{eqn:jeansstd}), relative to that from the generalized
      Jeans equation~(\ref{eqn:jeansgen3}).  The curves show the
      predictions using the parameters that we fit for
      \emph{halo 1} (black dot-dashed) and \emph{halo 2} (red
      dashed).}
\label{fig:rapporto}
\end{figure}

Instead of considering $\sigma_r$ for given mass and anisotropy profiles, one can estimate
the error in the mass profile derived from the standard Jeans
equation~(\ref{eqn:jeansstd}), relative to that derived from the generalized
Jeans equation, for given $\sigma_r(r)$ and $\beta(r)$.
Comparing with the standard and generalized Jeans equations, one finds
that the mass derived from the standard Jeans equation~(\ref{eqn:jeansstd})
can be written as
\begin{equation}
M_{\rm std-Jeans}(r) = \left [1 + {S(r)\over GM(r)/r^2} \right ]\,M(r) \ ,
\label{eqn:Mstd}
\end{equation}
where $M(r)$ is the mass profile obtained from the generalized Jeans
equation.
Fig.~\ref{fig:rapporto} shows that beyond the virial
radius, the corrections to the standard Jeans equation~(\ref{eqn:jeansstd})
are not negligible: in the range $\approx 2-4\,r_{\rm v}$, the correction causes the standard
Jeans equation to underestimate the total mass by $\approx 20-60\%$ for \emph{halo 1} and $\approx 20-40\%$ for \emph{halo 2}.

\section{Conclusions and Discussion}
The general purpose of this work is to improve our understanding of the
dynamics of galaxies that are still falling on to relaxed clusters, with the
motivation of performing a Jeans analysis of the mass profile out to several
virial radii, for possible cosmological applications.  The
standard Jeans equation describing the cluster dynamics, assumes the system
to be in equilibrium, with no mean radial streaming motions, 
and therefore it cannot be applied beyond the fully
virialized cluster zone. 

We have presented a generalized Jeans equation that takes into account the
non-zero mean radial velocities of galaxies outside the virial radius, as
well as the background density and the cosmological constant terms. We
accurately reproduce  the radial velocity
dispersion profiles of a  stack of 27 cluster-mass
haloes and of two isolated haloes, out to 4 virial radii.
In particular, while the standard Jeans equation provides accurate radial
velocity dispersions out to $\approx 2$ virial radii, it over estimates the
radial velocity dispersion by typically a factor 1.5 beyond $\approx
3$ virial radii.
In the standard Jeans formalism, the total mass is underestimated
  by $\approx\,20-60\%$ in the region $\approx 2-4\,r_{\rm v}$.

A consistent description of cluster dynamics in the infall region can be
 useful for an accurate dynamical mass measurement at the infall scale. 
The estimation of mass profiles with the standard Jeans
analysis involves the modelling of the line-of-sight velocity dispersion of
the tracer (i.e. galaxies in clusters) by solving the lowest-order Jeans equation, to compare with that obtained from taking the second moments of the observed galaxy
line of sight velocities.
This
also involves a measurement of the radial velocity anisotropy of galaxies.
Just as several approaches have been proposed to break the mass velocity
anisotropy degeneracy inherent to the standard Jeans equations, we wish to do
the same when applying the generalized Jeans equation all the way to several
virial radii, except that we also need to determine a third quantity :
the
mean infall velocity. 
As our parametrized approach to recover the radial velocity
dispersion profile beyond the virial radius requires 12 free
parameters, 
it should be viewed more as a proof of concept that the standard Jeans
equation is adequate for determining the radial velocity dispersion
profile up to 2 virial radii and inadequate beyond, rather than 
being a method to accurately determine the radial velocity dispersion
profile from observational data. 
\cite{Kara10} recover the infall pattern of the Virgo cluster, with the knowledge
of the depth along the line of sight obtained from
distance indicators independent of redshift \citep{Mei07}. However, 
 the uncertainties  on the velocities appear to be too large for
 reproducing 
 the radial velocity dispersion profile more accurately with the
 generalized Jeans equation than with
the standard Jeans equation.

The model also involves the logarithmic growth rate of the virial radius as
well as the departure from
self-similarity of the evolution of the infall velocity profile. We
have presented seven theoretical derivations for 
the logarithmic growth rate of the virial radius (as a function of mass or
concentration), which lead to similar values.
On the other hand, our stacked halo leads to a different non-self-similarity
parameter ($\alpha$) than our two isolated haloes. We suspect that this
parameter is not universal, but strongly depends on the mass accretion
history of the halo.
It would be useful to analyse the non-self-similarity parameter in more
detail with simulations.

We finally note that with infall present, the
kinetic energy is expected to be larger than in the case with no infall. The
virial ratio, $2K/W=1$, can be seen as a spatial integral over the Jeans
equation, where $K$ and $W$ represent the total kinetic energy and the total
potential energy of the system. One therefore expects the virial ratio to be
larger than unity for systems where infall is important \citep{CL96,PKK12}.

We are planning an extension of this work to test how far the standard Jeans
  equation is relevant in reproducing the line-of-sight velocity
  dispersion profile and possibly 
find signatures of infall in the shape of the line-of-sight velocity profile.

\section*{Acknowledgements} 
G.A.M. is indebted to James Binney and Ewa {\L}okas for useful discussions at
a very early stage of this work and Avishai Dekel for useful comments
throughout. He
thanks DARK for their hospitality during the visit that launched the
collaboration, while M.F. thanks the IAP for their hospitality during
two visits.
 The authors thank Antonio Cuesta for providing simulation data in
digital form, helping them to build Fig.~(\ref{fig:vpcuesta}).
The simulation has been performed at the Leibniz Rechenzentrum (LRZ) Munich.
The Dark Cosmology Centre is funded by the Danish National Research Foundation.

\bibliography{infall}

\onecolumn
\appendix

\section{Departure from self-similarity of the evolution of the peculiar
  velocity}

\label{sec:nonselfsim}
\begin{figure}
\centering
\includegraphics[width=0.5\hsize]{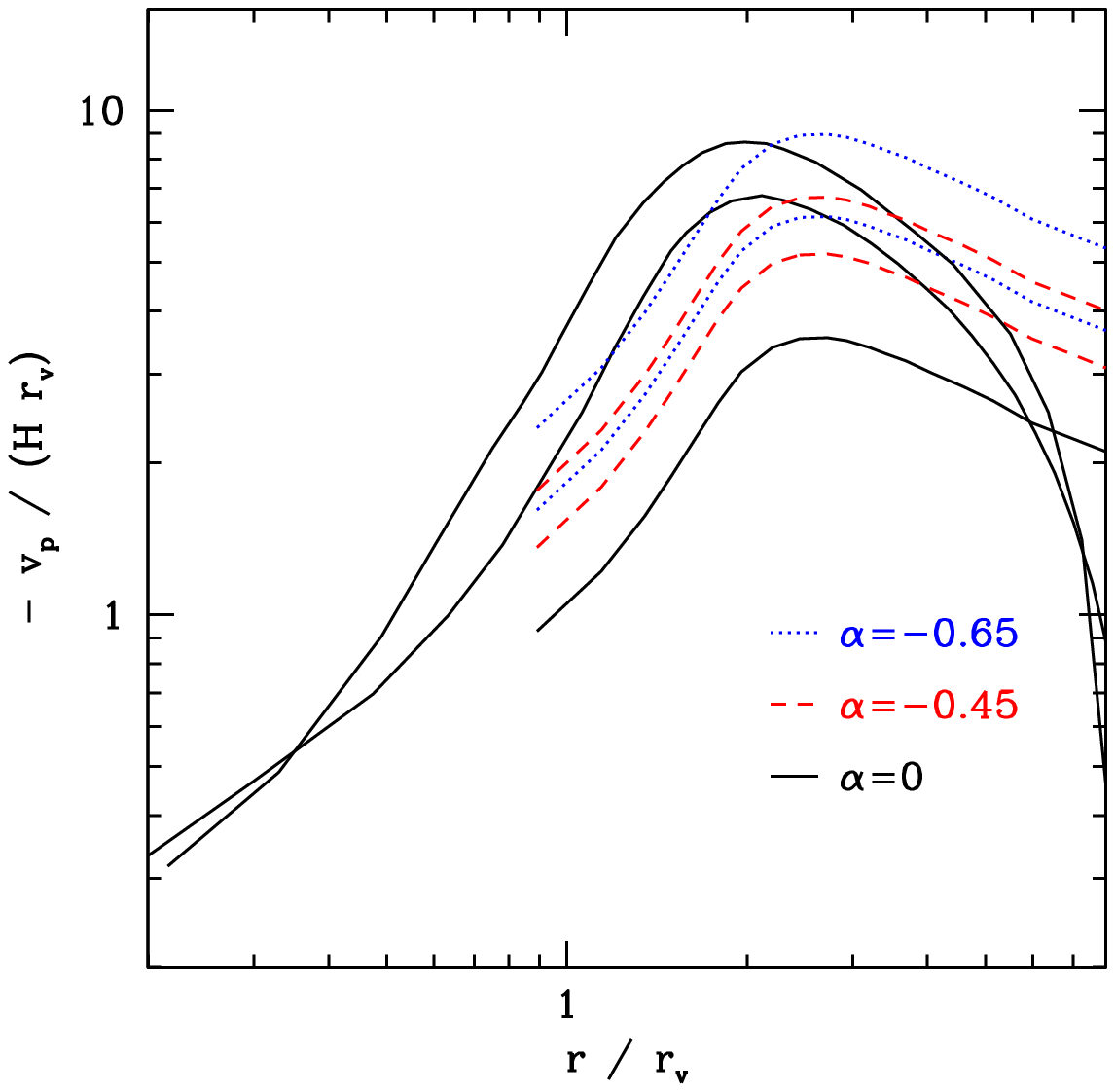} 
\caption{Time evolution of the mean peculiar velocity profile (in virial
  units).
The black curves show the peculiar
velocity measured in simulations at $z=0$, 1, and 2 (going upwards at $r =
2\,r_{\rm v}$), adapted from fig.~13c of  Cuesta et al. (2008).
The red and blue curves show the predictions for $z=1$ and 2 obtained by
extrapolating the $z=0$ curve with $(t/t_0)^\alpha$. Two values of $\alpha$
are shown: one ($\alpha=-0.65$) that matches best the maximum absolute value
of the peculiar
velocity and one ($\alpha=-0.45$) that represents a best compromise over the
relevant range of radii.}  
\label{fig:vpcuesta}
\end{figure}

\cite{CPKM08} have measured the mean radial infall velocity as a function
of radius, averaging over haloes of different mass, and repeating the exercise
at $z=0$, 1 and 2. Their fig.~13c shows that the infall pattern is not
self-similar, but instead decreases in time (in virial units). We converted
their mean radial velocity to peculiar velocity (subtracting the Hubble
flow). Fig.~\ref{fig:vpcuesta} shows that 
the evolution of the mean peculiar velocity in virial units is not
self-similar: the black curves do not lie on top of one another.
Assuming that $v_p(r,t)/[H(t) r_{\rm v}(t)]$ scales as $(t/t_0)^\alpha$ improves
the self-similarity. The peak peculiar velocities are reproduced for
$\alpha=-0.65$, but the absolute  
peculiar velocities at large radii are overpredicted.
Choosing $\alpha=-0.45$ represents a good compromise over the range of radii
that are relevant: those where we measure the radial velocity dispersion and
those slightly beyond that take part in the outwards integration of
equation~(\ref{eqn:jeansgen3}).

\section{Growth rate of the virial radius}
\label{sec:rvgrowth}

There are several ways to compute the logarithmic  growth rate of the virial
radius, ${\rm d}\ln r_{\rm v}/{\rm d}\ln t$.

\subsection{Self-similar growth in Einstein de Sitter universe}
\label{subsecb1}
For a single halo in a universe with present-day density and dark energy
parameters   
$\Omega_m=1$ and $\Omega_\Lambda=0$, one simply has
${\rm d}\ln r_{\rm v}/{\rm d}\ln t = 8/9 \simeq 0.89$, independent of the
halo mass \citep{GG72}.

\subsection{Exponential halo mass evolution with redshift}
\label{sec:wechsler}
\cite{Wechsler+02} analysed cosmological simulations and derived $M_{\rm v}(z) =
M(z=0)\,\exp(-\alpha z)$, with $\alpha \approx 0.6$ for haloes of mass close
to those considered here ($M \approx 8\times 10^{13} M_\odot$).
With
\begin{equation}
M_v =  {\Delta\over 2}\,{ H^2 \,r_{\rm v}^3\over G} \ ,
\label{Mrvir}
\end{equation}
this yields
\begin{equation}
{{\rm d}\ln r_{\rm v} \over {\rm d}\ln t} = {1\over 3} \,\left (
{{\rm d}\ln M_v\over {\rm d}\ln t} 
- {{\rm d}\ln \Delta \over {\rm d}\ln t} 
- 2 {{\rm d}\ln H\over {\rm d}\ln t} \right ) \ .
\label{dlnrdlnt}
\end{equation}
One can write ${\rm d}\ln M_{\rm v}/{\rm d}\ln t = -\alpha / ({\rm d}\ln t /
{\rm d}z)$.
Given that, for a flat universe,
one has (\citealp{Peebles93}, equation~13.20)
\begin{equation}
H_0 \,t(z) = {2\over 3\,\sqrt{1-\Omega_{\rm m}}}\,\sinh^{-1} \left
({\sqrt{(1/\Omega_{\rm m})-1}\over (1+z)^{3/2}}\right) \ ,
\end{equation}
one finds 
\begin{equation}
\left ({{\rm d}\ln t \over {\rm d}z}\right)_{z=0} = -{3\over
  2}\,{\sqrt{1-\Omega_{\rm m}}\over
  \sinh^{-1} \sqrt{1/\Omega_{\rm m}-1}} \ ,
\end{equation}
which tends to $-0.975$ for the density parameter of our simulation,
$\Omega_{\rm m}=0.24$.
We deduce that ${\rm d}\ln M_{\rm v} / {\rm d}\ln t \simeq 0.615$.

Given that, for a flat universe,
one also has 
(\citealp{Peebles93}, equation 13.3) 
\begin{equation}
E(z) = {H(z)\over H_0} = \sqrt{\Omega_{\rm m} (1+z)^3+1-\Omega_{\rm m}} \ ,
\label{Eofz}
\end{equation}
 one deduces
\begin{eqnarray} 
{{\rm d}\ln H\over {\rm d}\ln t} &\!\!\!\!=\!\!\!\!& 
-{\Omega_{\rm m} \over \sqrt{1-\Omega_{\rm m}}}\,(1+z)^3\,
{\sqrt{1-\Omega_{\rm m} \left [1-
  (1+z)^{3}\right]} 
\over 1+ \Omega_{\rm m}
  z  \left(3 + 3 z + z^2\right)} 
\sinh^{-1}\left({\sqrt{1/\Omega_{\rm m}-1}\over (1+z)^{3/2}}\right)
 \nonumber \\
&\!\!\!\!=\!\!\!\!& -{\Omega_{\rm m} \over 1-\Omega_{\rm m}}\, \sinh
  ^{-1}\left(\sqrt{1/\Omega_{\rm m}-1}\right)
\qquad (z=0) \ , \label{dlnHdlntz0} \\
&\!\!\!\!\simeq\!\!\!\!&  -0.402 - 
 1.08\, (\Omega_{\rm m}\!-\!0.27) 
- [0.638 + 0.698\, (\Omega_{\rm m}\!-\!0.27)]\, z \ ,
\end{eqnarray} 
where the approximation is from a series expansion and is accurate to better
than 1.7\% (0.7\% rms) for $0<z<0.2$ and $0.24 < \Omega_{\rm m} < 0.3$.
So for our case of $\Omega_{\rm m}=0.24$ and $z=0$, equation~(\ref{dlnHdlntz0}) 
yields ${\rm d}\ln H/{\rm d}\ln t\simeq -0.37$.

Finally, 
using the approximation \citep{BN98}
\begin{equation}
\Delta \simeq 18\,\pi^2 + 82\,[\Omega_m(z)-1] - 39\,[\Omega_m(z)-1]^2 
\end{equation}
where
\begin{equation}
\Omega_{\rm m}(z) = {\Omega_{\rm m}\,(1+z)^3\over E^2(z)}
\end{equation}
(see eq.~[\ref{Eofz}]), we find the series expansion
\begin{equation}
{{\rm d}\ln \Delta \over {\rm d}\ln t} \simeq
-0.841 + 0.570\,(\Omega_{\rm m}-0.27)  + [0.337 + 5.47\, (\Omega_{\rm m}-0.27)]\, z +
[1.447-6.01\,(\Omega_{\rm m}-0.27)]\,z^2 \ ,
\label{dlnDeltadlnt}
\end{equation}
which is accurate to better than 2.4\% (0.9\% rms) for $0<z<0.2$ and
$0.24 < \Omega_{\rm m} < 0.3$.
For $\Omega_{\rm m}=0.24$ and $z=0$,
equation~(\ref{dlnDeltadlnt}) yields ${\rm d}\ln \Delta / {\rm d}\ln t\simeq
-0.85$.
Putting this altogether, we deduce that ${\rm d}\ln r_{\rm v}/{\rm d}\ln t
\simeq 0.735$.

\subsection{Scaling with inverse Hubble time}
\cite{ZMJB03} noted that $r_{\rm v} \propto 1/H$. With ${\rm d}\ln H/{\rm
  d}\ln t\simeq -0.37$ (Section~\ref{sec:wechsler}), we obtain 
${\rm d}\ln r_{\rm v}/{\rm d}\ln t\simeq 0.37$ for
the \citeauthor{ZMJB03} approximation.

\subsection{Constant circular velocity}
\cite{MTTC12} noted that the mean growth of haloes follows roughly $v_{\rm
  circ}(r_{\rm v})= \rm const$. 
Equation~(\ref{Mrvir}) then implies that $r_{\rm v} \propto 1/(H \sqrt{\Delta})$.
With  ${\rm d}\ln H/{\rm  d}\ln t\simeq -0.37$ and 
 ${\rm d}\ln \Delta / {\rm d}\ln t\simeq
-0.85$
 (Section~\ref{sec:wechsler}), we derive
${\rm d}\ln r_{\rm v}/{\rm d}\ln t\simeq 0.795$ for the constant circular
velocity approximation,
close to the $\Omega_{\rm m}=1$ slope, but far
from the slope with the \citeauthor{ZMJB03} approximation.

\subsection{Halo merger rate in Millennium simulations}
\cite{FMB10} have measured the halo merger rate in the Millennium and
Millennium-II cosmological dark matter simulations. Their equation~(2) provides
the mean and median mass growth rates as ${\rm d} M_v/dt = a\,(M_v/10^{12} M_\odot)^{1.1}
(1+b\,z) E(z)$, with $a=46$ (mean) or 25 (median) $M_\odot\,\rm yr^{-1}$ and $b=1.11$ (mean)
or 1.65 (median). Hence, 
\begin{equation}
{{\rm d}\ln M_v \over {\rm d}\ln t} = {a \over 10^{12}} \,t(z)\,
\left ({M_v \over 10^{12} M_\odot}\right)^{0.1} (1+b\,z)\, E(z) \ ,
\label{dlnMdlntFMB}
\end{equation}
where $t(z)$ is measured in yr.
The (slightly) positive slope on
mass recovers the fact that high-mass haloes are rare today and even rarer in
the past, and must therefore grow faster.
Combining with equation~(\ref{Mrvir}), one obtains
\begin{eqnarray}
{{\rm d}\ln r_{\rm v} \over {\rm d}\ln t} &\simeq&
0.548  + 0.00511\,{a\over h} + \left (0.530-0.005\,{a\over h}\right)\,
(\Omega_{\rm m}\!-\!0.27) \nonumber \\
&\mbox{}& \qquad  
+ \left [0.426 - 0.00308 \,{a\over h}\,(1-1.66\,b) \right]\,z +
0.00118\,{a\over h}\,(\log M\!-\!14)
\ .
\label{dlnrdlntFMBser}
\end{eqnarray}
Equation~(\ref{dlnrdlntFMBser}) is accurate to 4\% (1.4\% rms) for
$0< z < 0.2$, $0.24 < \Omega_m < 0.3$, and $12 < \log M/M_\odot < 15.4$. 
Equations~(\ref{dlnrdlnt}) and (\ref{dlnMdlntFMB}) 
yield ${\rm d}\ln r_{\rm v}/{\rm d}\ln t \simeq 0.74$, 0.79, 0.86, and 0.95
(mean) or 0.64, 0.67, 0.71 and 0.76 (median)
for $\Omega_{\rm m}=0.24$, $h=0.73$, $z=0$, and $\log M$ = 12, 13, 14, and 15, respectively.

\subsection{Extended Press-Schechter theory}

\cite{ND08a} use extended Press-Schechter theory to derive a mass growth
rate that can be written as ${\rm d}\ln M_v / {\rm d}\ln t = -\alpha\, t \,\dot \omega \,(M/10^{12}
M_\odot)^\beta$ with $\dot \omega \simeq
-0.047\,[1+z+0.1\,(1+z)^{-1.25}]^{2.5}\,(h/0.73)\,\rm Gyr^{-1}$, $\alpha=0.59$ and
$\beta=0.141$.
With equations~(\ref{Eofz}), (\ref{dlnrdlnt}), and (\ref{dlnDeltadlnt}), this
leads to the series expansion
\begin{eqnarray}
{{\rm d}\ln r_{\rm v} \over {\rm d}\ln t} &\simeq& 0.548 + 0.264 \,\alpha \,10^{2\beta} 
+ \left(0.530 - 0.268 \,\alpha\,10^{2\beta}\right )\,(\Omega_m-0.27) \nonumber
\\
&\mbox{}& \qquad + \left (0.426+0.259 \,\alpha\,10^{2\beta}\right)\,z +
0.607\,\alpha\,\beta\,10^{2\beta}\,(\log M\!-\!14) \ .
\label{dlnrvdlntND}
\end{eqnarray}
Equation~(\ref{dlnrvdlntND}) is good to
7.6\% (2.4\% rms)  accuracy in the range $0<z<0.2,0.24<\Omega_{\rm m}<0.3,12<\log
M/M_\odot<15.4$. 
The exact solution
yields ${\rm d}\ln r_{\rm v}/ {\rm d}\ln t = 0.69$, 0.75, 0.84, and 0.96 for
$z=0$, $\Omega_{\rm m}=0.24$, and $\log M = 12$, 13, 14 and 15, respectively.

\subsection{Minimum growth rate}
We can estimate a \emph{minimum} growth rate by considering the growth of a
single halo in a uniform universe.
Assuming an NFW density profile at all times, with mass profile
\begin{eqnarray}
M(r,t) &=& M(a)\,\widetilde M\left ({r\over a}\right) \\
\widetilde M\left ({r\over a}\right) &=& {\ln (x+1)-x/(x+1) \over \ln
  2-1/2} \ ,
\label{mtilde}
\end{eqnarray}
where $a$ is the radius of slope $-2$ and does not vary in time.
The virial radius $r_{\rm v}$ is the solution to
\begin{equation}
{3\,M(r,t) / \left ( 4 \pi\, r^3 \right) \over 3\,H^2(t) /
  8\pi \,G} = \Delta(t)
\end{equation} 
i.e., using equation~(\ref{mtilde}),
\begin{equation}
{2\,G\,M(a)\over a^3}\,{\widetilde M(c)\over c^3} = \Delta(t) \,H^2(t)\ ,
\label{meandens0}
\end{equation}
where $c=r_{\rm v}/a$ is the concentration parameter. Now we do not need to solve
equation~(\ref{meandens0}) for $c$ to obtain the growth rate of $r_{\rm v}$.
Indeed, at time $t+dt$, where $dt\ll t$, equation~(\ref{meandens0}) becomes
\begin{equation}
{2\,G\,M(a)\over a^3}\,{\widetilde M(c)\over c^3}\, 
\left [1+ \left({{\rm d}\ln \widetilde M\over {\rm d}\ln x}\right)_{x=c}\,{dc\over c}-3\,{dc
\over c} \right ] =  \Delta(t)\,H^2(t) + {{\rm d}\left(\Delta\,H^2\right)\over {\rm d}\ln t}\,{\rm d}\ln t \ .
\label{meandens1}
\end{equation}
Dividing equation~(\ref{meandens1}) by equation~(\ref{meandens0}),
one obtains
\begin{equation}
{{\rm d}\ln r_{\rm v}\over {\rm d}\ln t} = {{\rm d}\ln c\over {\rm d}\ln t} = {({\rm d}\ln \Delta / {\rm d}\ln t) +
  2\,({\rm d}\ln H/{\rm d}\ln t) \over
  ({\rm d}\ln \widetilde M / {\rm d}\ln x)_{x=c} - 3} \ .
\label{dlnrdlntmin}
\end{equation}
After a series expansion, equation~(\ref{dlnrdlntmin}) becomes
\begin{equation}
{{\rm d}\ln r_{\rm v}\over {\rm d}\ln t} \simeq 
0.723  +0.413\,z + 0.698\,(\Omega_{\rm m}-0.27) - 0.207\,(\log c -0.7)  -
1.79\,(\Omega_{\rm m}-0.27)\,z \ .
\label{dlnrdlntminser}
\end{equation}
The approximation of equation~(\ref{dlnrdlntminser}) is accurate to 3.7\%
(1.4\% rms) for $0<z<0.2$, $0.24<\Omega_{\rm m}<0.3$, and $0.5<\log c < 1$.
For $\Omega_{\rm m}=0.24$ and $z=0$, equation~(\ref{dlnrdlntmin}) yields
${\rm d}\ln r_{\rm v}/{\rm d}\ln t = 0.75$, 0.70, 0.67, and 0.65 for $c=3$, 5, 7, and 10,
respectively.
One therefore notices that \citeauthor{ZMJB03}'s approximation of $r_{\rm v}
\propto 1/H$ yields a slower growth rate for $r_{\rm v}$ than our minimum growth
rate found here.
Note that, for $\Omega_{\rm m}=1$ and $z=0$, the minimum growth rate is
${\rm d}\ln r_{\rm v}/{\rm d}\ln t \simeq 0.879 + 0.252\,(\log c -
0.7)$, not far from the Einstein de Sitter universe growth rate
(\citealp{GG72}, see Section~\ref{subsecb1}), with equality for $c=4.57$.

\subsection{Summary}
In summary, for the cosmology of our simulation, at $z=0$ for our mean halo
mass of $8.6\times 10^{13} M_\odot$ and concentration parameter  $c=6.4$, we
find
${\rm d}\ln r_{\rm v}/{\rm d}\ln t = 0.86$ (from \citeauthor{FMB10}'s
analysis of the merger rate in the Millennium simulations),
0.795 (for our constant circular velocity approximation), 0.735 (from
\citeauthor{Wechsler+02} et al's exponential mass growth),
0.71 (from \citeauthor{ND08a}'s extended Press-Schechter theory),
0.68 for the minimal growth scenario, but only 0.37 for \citeauthor{ZMJB03}'s
scaling with
inverse Hubble time. We consider this last scaling as inaccurate and we 
adopt ${\rm d}\ln r_{\rm v}/{\rm d}\ln t = 0.7$, slightly above our minimal
growth scenario.

\section{Relations between slopes}
\label{sec:slopes}
Various numerical simulations have identified a range of apparent
universalities, where some are identified in cosmological simulations, and
others are found in controlled simulations. Most of these universalities are
usually considered in radial ranges where the systems are fully equilibrated. Since
we are considering here radial ranges much beyond the virial radius, it is
relevant to study these properties at large radii.

\begin{figure}
	\centering
	\includegraphics[angle=0,width=0.5\hsize]{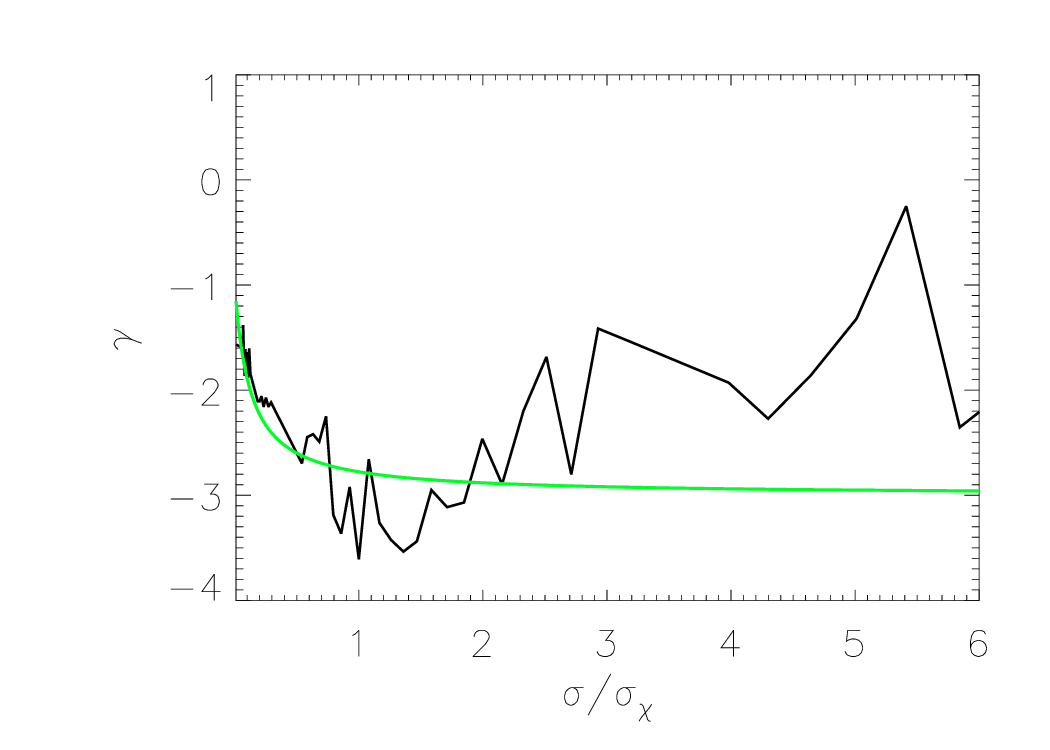}
	\caption{ Logarithmic derivative of the density profile
          $\gamma$ (equation~\ref{eqn:g}). Starting from $\gamma\approx-1$,
          the slope of the density function decreases reaching
          $\approx-3.2$ at around two times the virial radius, and in the outer region
          it increases again approaching zero. The green line shows the
          prediction for the double slope profile with parameters in
          Table~~\ref{par2}}.
\label{fig:gamma}
\end{figure}

Probably the most famous universality is the density profile as a function of
mass \citep{NFW96}. It suggests that the density slope
\begin{equation}
\label{eqn:g}
\gamma=\frac{{\rm d}\log\rho}{{\rm d}\log r}
\end{equation}
 has a smooth transition from $-1$ in the inner region, to $-3$ in the outer region. 
The corresponding plot is shown for the stacked clusters in
Fig.~\ref{fig:gamma}. It is clearly seen that around the virial
radius the density profile flattens out, and the logarithmic slope
approaches $-1$ around three times the virial radius.
The green line shows the prediction for the double slope profile with
parameters in Table~\ref{par2}.

\begin{figure}
	\centering
	\includegraphics[angle=0,width=0.5\hsize]{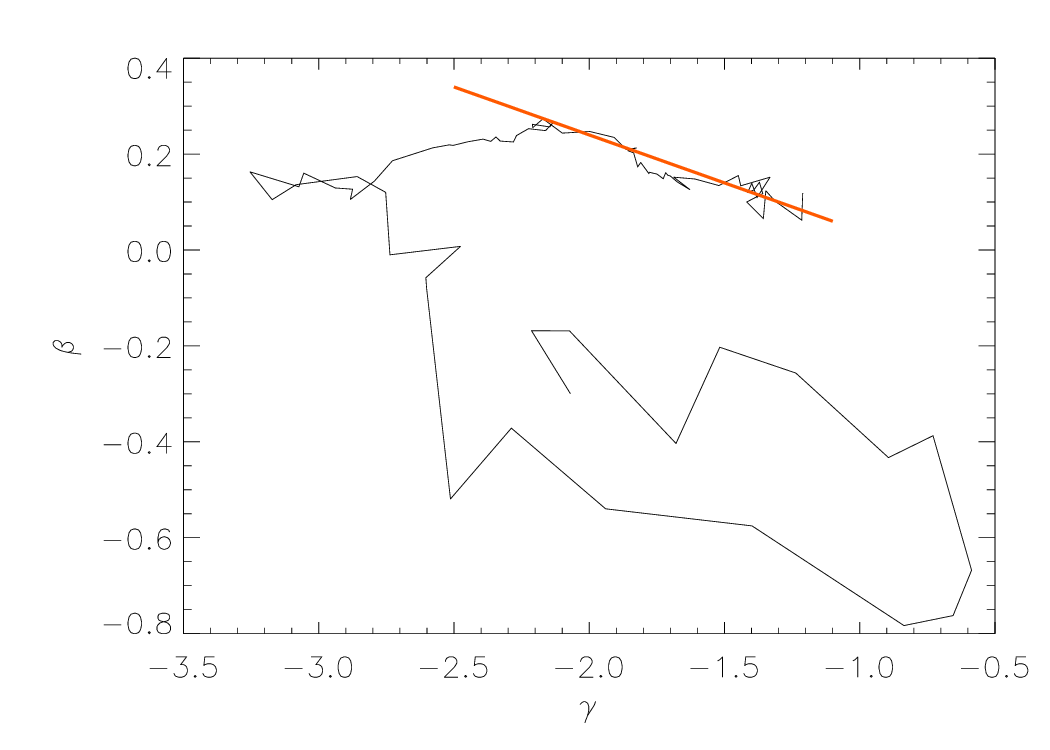}
	\caption{Velocity anisotropy $\beta$ as a function of the
          radial derivative of the density profile $\gamma$. The red solid line corresponds to equation~(\ref{eqn:han}) \citep{HS2003}.}
\label{fig:betagamma}
\end{figure}

Fig.~\ref{fig:betagamma} shows the relation between $\gamma$ and the  velocity dispersion anisotropy given by equation~(\ref{eqn:beta}). This is quite in agreement with the universality proposed in \citep{HS2003}
\begin{equation}
\label{eqn:han}
\beta=-0.2\,(\gamma+0.8)
\end{equation}
 in the inner region, but departs significantly for $\gamma<-2.2$.

\begin{figure}
	\centering
	\includegraphics[angle=0,width=0.5\hsize]{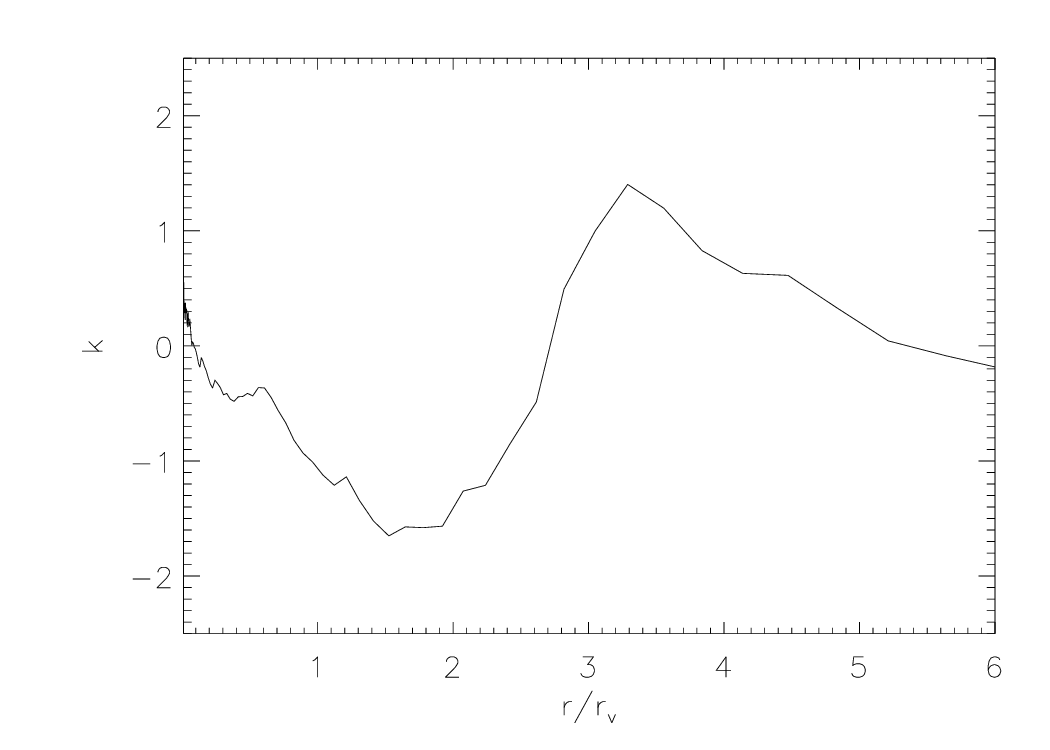}
	\caption{Logarithmic derivative of the radial velocity dispersion $\kappa$, as defined by equation~\ref{eqn:k}.}
\label{fig:kappa}
\end{figure}

\begin{figure}
	\centering
	\includegraphics[angle=0,width=0.5\hsize]{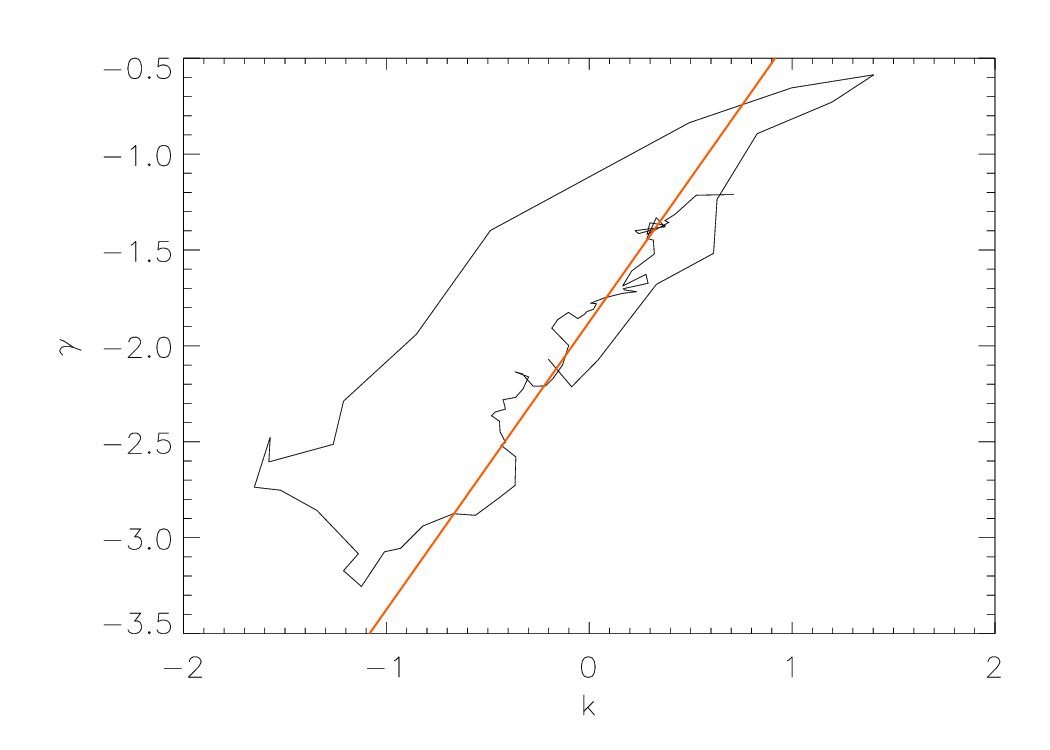}
	\caption{Derivative of the radial velocity dispersion $\kappa$ as a function of the logarithmic slope of the density profile $\gamma$. The red solid line corresponds to the equation~(\ref{eqn:tn2}) \citep{TN01, Ludlow+11}.}
\label{fig:kappagamma}
\end{figure}

A connection between $\gamma$ and the radial velocity dispersion was suggested by \citep{TN01, Ludlow+11}
\begin{equation}
\label{eqn:tn}
\frac{\rho}{\sigma^{3}}\approx r^{-\alpha},
\end{equation}
where $\alpha=1.875$, in agreement with the prediction from the spherical infall model \citep{Bertschinger85}.
In Fig.~\ref{fig:kappa} the derivative of the velocity dispersion is displayed:
\begin{equation}
\label{eqn:k}
\kappa=\frac{{\rm d}\log\sigma_r^{2}}{{\rm d}\log r}.
\end{equation}
 Equation~(\ref{eqn:tn}) in terms of $\kappa$ and $\gamma$ reads
\begin{equation}
\label{eqn:tn2}
\gamma=-\alpha+\frac{3}{2}\kappa \ .
\end{equation}
The connection in the $\gamma-\kappa$ space is in fair agreement with the formula (\ref{eqn:tn2}), as we show in Fig.~\ref{fig:kappagamma}.

\end{document}